\documentclass[aps,prl,showpacs,twocolumn,superscriptaddress,letterpaper]{revtex4}
\usepackage[normalem]{ulem}
\usepackage[dvips]{color}
\usepackage{graphicx}
\usepackage{dcolumn}
\usepackage{bm}
\usepackage{amsmath, amssymb}%
\usepackage{epstopdf}
\usepackage{multirow}
\usepackage{amsmath}

\newcommand\Esym{\ifmmode E_{sym}(\rho) \else $E_{sym}(\rho)$\fi}
\newcommand\EsymZ{\ifmmode E_{sym}(\rho_0) \else $E_{sym}(\rho_0)$\fi}
\newcommand\EsyK{\ifmmode E_{sym}^{kin} (\rho) \else $E_{sym}^{kin} (\rho)$\fi}
\newcommand\EsyKZ{\ifmmode E_{sym}^{kin} (\rho_0) \else $E_{sym}^{kin} (\rho_0)$\fi}
\newcommand\EsyP{\ifmmode E_{sym}^{pot} (\rho) \else $E_{sym}^{pot} (\rho)$\fi}
\newcommand\EsyPZ{\ifmmode E_{sym}^{pot} (\rho_0) \else $E_{sym}^{pot} (\rho_0)$\fi}

\renewcommand\sout{\bgroup \color{red} \ULdepth=-.5ex \ULset}

\begin{document}

\title{Symmetry energy of nucleonic matter with tensor correlations}

\newcommand*{\TAU }{School of Physics and Astronomy, Tel Aviv University, Tel Aviv 69978, Israel}
\newcommand*{\TAUindex}{1}
\affiliation{\TAU} 

\newcommand*{\Texas }{Department of Physics and Astronomy, Texas A$\&$M University-Commerce, Commerce, Texas 75429-3011, USA}
\newcommand*{\Texasindex}{2}
\affiliation{\Texas} 

\newcommand*{\China }{College of Science, University of Shanghai for Science and Technology, Shanghai, 200093, China}
\newcommand*{\Chinaindex}{3}
\affiliation{\China} 

\newcommand*{\ODU }{ Old Dominion University, Norfolk, Virginia 23529,USA} 
\newcommand*{\ODUindex}{4}
\affiliation{\ODU }

  
\author{Or Hen}
\email[Contact Author \ ]{or.chen@mail.huji.ac.il}
     \affiliation{\TAU}
\author{Bao-An Li}
\affiliation{\Texas}
\author{Wen-Jun Guo}
\affiliation{\Texas}\affiliation{\China}
\author{L.B. Weinstein}
     \affiliation{\ODU}
\author{Eli Piasetzky}
\affiliation{\TAU}

\date{\today}

\begin{abstract} 
The nuclear symmetry energy ($E_{sym}(\rho)$) is a vital ingredient of our understanding of many processes, from heavy-ion collisions to neutron stars structure.  While the total nuclear symmetry energy at nuclear saturation density ($\rho_0$) is relatively well determined, its value at supranuclear densities is not. The latter can be better constrained by separately examining its kinetic and potential terms and their density dependencies.  The kinetic term of the symmetry energy, $\EsyKZ$, equals the difference in the per-nucleon kinetic energy between pure neutron matter (PNM) and symmetric nuclear matter (SNM), often calculated using a simple Fermi gas model. However, experiments show that tensor force induced short-range correlations (SRC) between proton-neutron pairs shift nucleons to high-momentum in SNM, but have almost no effect in PNM. We present an approximate analytical expression for $E_{sym}^{kin}(\rho)$ of correlated nucleonic matter.   In our  model,  $E_{sym}^{kin}(\rho_0)=-10$ MeV, which differs significantly from  $+12.5$ MeV for the widely-used free Fermi gas model.  This result is consistent with our analysis of recent data on the free proton-to-neutron ratios measured in intermediate energy nucleus-nucleus collisions as well as with microscopic many-body calculations, and previous phenomenological extractions.
We then use our calculated $E_{sym}^{kin}(\rho)$  in combination with the known total symmetry energy and its density dependence at saturation density to constrain the value and density dependence of the potential part and to extrapolate the total symmetry energy to supranuclear densities. \end{abstract} 


\maketitle

The nuclear symmetry energy, $E_{sym}(\rho)$, where $\rho$ is the nuclear density, is related to the difference in the energy per nucleon of pure neutron matter (PNM) and symmetric nuclear matter (SNM).  
It determines many nuclear and astrophysical properties, such as the cooling of proto-neutron stars \cite{Lat91}, the mass-radius relations of neutron stars \cite{Pra88b}, properties of nuclei involved in $r$-process nucleosynthesis~\cite{Nikolov11}, and heavy-ion collisions \cite{LiBA97a,Che05a,Tsang09}.

Much effort is being invested in improving our knowledge of \Esym.  In particular, several major radioactive beam facilities being built around the world have all listed constraining the symmetry energy as one of their major science drivers, see, e.g., Ref.~\cite{FRIB}.  
Moreover, observations of neutron stars from current missions such as the Chandra X-ray and XMM-Newton observatories, and upcoming missions such as the Neutron star Interior Composition ExploreR (NICER) \cite{NASA} will provide high precision data to infer more accurately neutron star radii which are very sensitive to the symmetry energy \cite{Andrew1,LiBA06a,Will,Andrew} 

Significant progress has been made in recent years in constraining \Esym{} especially around $\rho\approx\rho_0$, the nuclear saturation density, using data from both terrestrial laboratory experiments and astrophysical observations \cite{Lynch09,Trau12,Tsang12,EPJA,Hor14,Jim13}.  Recent surveys of model analyses of world data found that the mean values of the symmetry energy and its density dependence at $\rho_0$ are consistent with $29\leq E_{sym}(\rho_0)\leq33$ MeV and $40\leq L=3\rho\frac{\partial E_{sym}(\rho)}{\partial \rho}|_{\rho_{0}}\leq60$ MeV~\cite{LiBA13, Lat13}. However, the decomposition of the symmetry energy into its kinetic and potential parts and its behavior at both sub-saturation ($\rho<\rho_0$) and supra-saturation ($\rho>\rho_0$) densities are still poorly known.

A common method to improving our knowledge of
the total symmetry energy,  \Esym{}, is to separate it into its potential ($E_{sym}^{kin} (\rho)$) and kinetic ($E_{sym}^{pot} (\rho)$) parts,
\begin{equation} \label{eq:a}
E_{sym}(\rho)  = E_{sym}^{kin} (\rho) + E_{sym}^{pot} (\rho) 
\end{equation}
and probing them separately~\cite{Andrew,Tsang09,Jim13}. The kinetic part of the symmetry energy, \EsyK, can be readily calculated from the nuclear momentum distribution.   The much less well understood potential part can then be calculated as $\EsyP=\Esym-\EsyK$.

This separation is valuable for several reasons.  As \EsyK{} and \EsyP{} have different density dependencies (typically parametrized as $E_{sym}^{kin} (\rho_0) (\frac{\rho}{\rho_0})^{\alpha}$ and $E_{sym}^{pot}(\rho_0)(\frac{\rho}{\rho_0})^{\gamma}$)
the total symmetry energy can be more reliably extrapolated to higher densities by extrapolating its kinetic and potential parts separately.
Secondly, knowledge of \EsyP{} is important for constraining key parameters in calculations of the symmetry energy, such as three-body forces \cite{Gandolfi14} and high-order chiral effective interactions~\cite{Hebeler14}.  These improved models then allow extrapolation of \EsyP{} to supra-saturation densities with improved accuracy \cite{Jim13,Steiner12,Hebeler13,Gezerlis13}.  Thirdly, knowing \EsyK{} and \EsyP{} separately is required to describe heavy-ion reactions and describe the isovector dynamical observables. For example, the density dependence of \Esym{} as extracted from heavy-ion collisions depends on models of \EsyK{} \cite{Li14}.


The kinetic part is often approximated in a nonrelativistic free Fermi gas model~\cite{Tsang09,Andrew} as the per-nucleon difference between the kinetic energy of pure neutron matter at a density $\rho$ and the kinetic energy of symmetric nuclear matter where the protons and neutrons each have density $\rho/2$:
\begin{equation}
\EsyK\vert_\textrm{FG} = (2^{\frac{2}{3}}-1)\frac{3}{5}E_F(\rho)\approx 12.5 \hbox{MeV} (\rho/\rho_0)^{2/3}
\end{equation}
where $E_F(\rho)$ is the Fermi energy at density $\rho$. 

However, short-range correlations (SRC) due to the tensor force acting predominantly between neutron-proton pairs significantly increase the average momentum and hence the kinetic energy in SNM but have almost no effect in PNM.  They thus reduce significantly the kinetic symmetry energy, possibly even to negative values.  This has been shown recently in both phenomenological models \cite{XuLi} and microscopic many-body theories \cite{Vid11,Lov11,Car12,Rio14}.  
At fixed symmetry energy, \Esym, the SRC induced decrease of \EsyK{} increases \EsyP{} beyond its Fermi Gas model limit of $\EsyPZ = \EsymZ - \EsyKZ\vert_{FG} \approx 19.1$ MeV.
This is important for transport model simulations of heavy-ion collisions \cite{LiBA97a,Che05a,Tsang09,Bar05,LCK}.

In this paper we provide a phenomenological analytical expression for the kinetic symmetry energy of correlated nucleonic matter based on calculations of nuclear momentum distributions and on data at saturation density ($\rho_0$) from inclusive $(e,e^{\prime})$ and exclusive $(e,e'pN)$ scattering experiments at the Thomas  Jefferson National Accelerator Facility (JLab) \cite{Egi03,Egi06,Fom12,Hen14a,sub}.  We give credence to our model by comparing to a transport model analysis of nucleon emission data in intermediate energy heavy-ion collisions \cite{MSU1,MSU2} and to many-body theoretical calculations of nuclei  and nuclear matter~\cite{Ciofi96,Wiringa14,Vid11,Lov11,Car12,Rio14}. Last we use the known values of the total symmetry energy, \Esym, and its density dependence, $L$, at saturation density to extract the total symmetry energy at supranuclear densities and to constrain the value and density dependence of the potential part of the symmetry energy.

It has long been known that the tensor force induced SRC leads to a high-momentum tail in the single-nucleon momentum distribution around 300--600 MeV/c~\cite{Bethe71,ant}. This high-momentum tail scales, i.e., its shape is almost identical for all nuclei from deuteron to infinite nuclear matter, see, e.g. Refs. \cite{fan,Pie92,Ciofi96}. This is shown by the constancy of the ratio of the per nucleon inclusive $(e,e^{\prime})$ cross sections for nucleus $A$ to the deuteron, $a_2(A)$, for Bjorken scaling parameter $x_B$ between about $1.5$ and $1.9$ \cite{Day89,Egi03,Egi06,Fom12}.    The ratio of the momentum distribution in nucleus $A$ to the deuteron for $300 \le k \le 600 $ MeV/c is just the cross section ratio $a_2(A)$. Extrapolation of the measured $a_2(A)$ to infinite SNM using three different techniques \cite{Cio91,Sar11,Pia13}, yield an average value of $a_2(\infty) = 7\pm1$.  The uncertainty in the extrapolation represents about 50\% of the difference between $a_2(A)\approx 5$ for heavy nuclei and $a_2(\infty)=7$ for SNM.

Exclusive two-nucleon knockout experiments \cite{pia,sub,Tan03,Bag10,Hen14a} show that, for $300\le k \le 600$ MeV/c,  proton knockout is accompanied by a recoil second nucleon and that second nucleon is predominantly a neutron, i.e., that $np$-SRC pairs dominate over $pp$ pairs by a factor of about 20. For recent reviews, see Refs.~\cite{fra,Arr12}.  This implies that correlations are about 20 times smaller in PNM than in SNM.
Since the integral of the deuteron momentum distribution from 300 to 600 MeV/c is about 4\% \cite{Pas02} and $a_2(\infty)= 7\pm 1$, the probability to find a high-momentum nucleon in SNM is about 28\% and in PNM is about $1-2$\%. 

The deuteron momentum distribution, $n_d(k)$, decreases as $1/k^4$ for $300 \le k \le 600$ MeV/c \cite{Hen14b}.  Since the nuclear momentum distribution, $n_A(k)$, in that range is predominantly due to $np$-SRC pairs and since it is proportional to the deuteron distribution, we can write that $n_A(k/k_F)(k/k_F)^4=R_d a_2(A)$, where $R_d=0.64\pm 0.10$ is extracted from the deuteron momentum distribution, and $k_F$ is the Fermi momentum \cite{Hen14b}.  At higher momenta, the momentum distribution $n(k)$ drops much more rapidly.

 This is supported by ``exact'' variational Monte Carlo (VMC) momentum distributions calculated \cite{Wiringa14} for $^4$He and
 $^{10}$B which decrease as $k^{-4}$ for $np$ pairs with small pair
 center-of-mass momentum for nucleon
 momenta $1.2 < k/k_F < 3$ to within about 10\%.  

\begin{figure} [b]
\includegraphics[width=\linewidth]{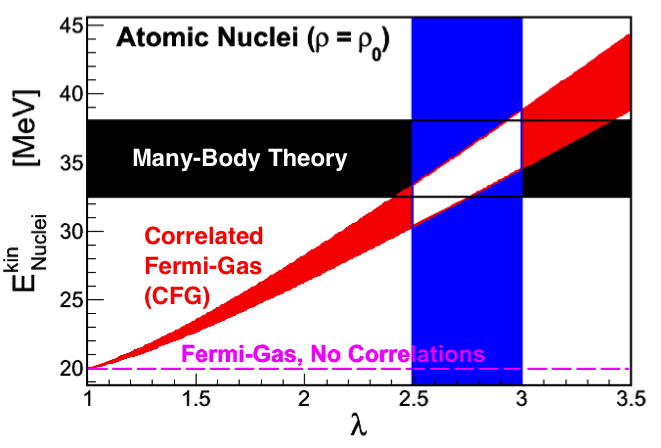}
\caption{(Color online) The per-nucleon kinetic energy  calculated using the Correlated Fermi Gas (CFG) model (diagonal red band)  for atomic nuclei from $^{12}$C to $^{208}$Pb. The calculated kinetic energy is shown as a function of $\lambda$, the high-momentum tail cutoff parameter. The vertical blue band shows the constraints on $\lambda$ from the deuteron momentum distribution. The diagonal red band reflects the model uncertainties.  Also shown are the results from the uncorrelated Fermi Gas model (dashed purple line) and a horizontal black band spanning the results from many-body nuclear calculations for various nuclei from $^{12}$C to $^{208}$Pb~\cite{Ciofi96} and from exact variational Monte Carlo (VMC) calculations for $^{12}$C~\cite{Wiringa14}.}
\label{fig:ekin}
\end{figure}

We therefore model $n(k)$ for SNM with a depleted Fermi gas region and a correlated high-momentum tail:
\begin{equation}
n^{SRC}_{SNM}(k)=\left\{ \begin{array}{ll}
A_0 & k<k_F\\
C_{\infty}/k^4 & k_F<k<\lambda k^0_F\\
0 & k>\lambda k^0_F
\end{array}
\right. 
\label{eq:nk}
\end{equation}
where $C_{\infty}=R_d a_2(\infty) k_F\equiv c_0 k_F$ is the phenomenological height factor~\cite{Hen14b}, $c_0=4.16\pm0.95$, $k_F^0$ is the Fermi momentum at $\rho_0$ and $\lambda\approx 2.75\pm 0.25$ is the high-momentum cutoff obtained from the momentum distribution of the deuteron \cite{Hen14b}. 
$A_0$ is a constant given by
\begin{equation}
A_0=\frac{3\pi^2}{(k^0_F)^3}\frac{\rho_0}{\rho}\left[1-\bigl[1-\frac{1}{\lambda}(\frac{\rho}{\rho_0})^{1/3}\bigr]\frac{c_0}{\pi^2}\right],
\end{equation}
determined by the normalization 
\begin{equation}
\frac{4\pi}{(2\pi)^3}\int^{\lambda k^0_F}_{0} 2 n^{SRC}_{SNM}(k) k^2 dk\equiv 1.
\end{equation}
Based on the JLab data \cite{sub}, fewer than 2\% of neutrons belong to $nn$-SRC pairs. We thus use the free Fermi gas model for PNM and include the 2\% upper limit for correlated neutrons in our estimate of the uncertainty band. 
In what follows we refer to this as the Correlated Fermi Gas (CFG) model.

\begin{figure} [b]
\includegraphics[width=\linewidth]{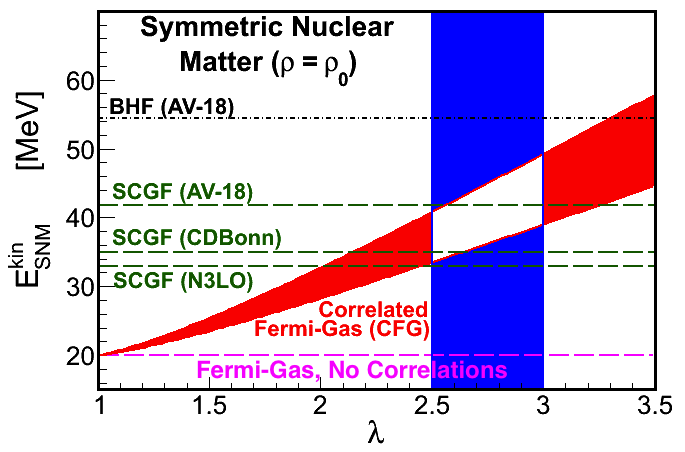}
\caption{(Color online) The per-nucleon kinetic energy for symmetric nuclear matter calculated using the Correlated Fermi Gas (CFG) model (red band). The calculated kinetic energy is shown as a function of $\lambda$, the high-momentum tail cutoff parameter. The blue band shows the constraints on $\lambda$ from the deuteron momentum distribution. The red band reflects the model uncertainties.  Also shown are the results from the uncorrelated Fermi Gas model (dashed purple line), the Brueckner-Hartree-Fock (BHF) model using the AV-18 interaction \cite{Vid11}, and the Self-Consistent Greens Function (SCGF) approach using the CDBonn, N3LO, and AV18  nucleon-nucleon interactions \cite{Car12,Rio14}.}
\label{fig:ekin_SNM}
\end{figure}

\begin{figure} [t]
\includegraphics[width=\linewidth]{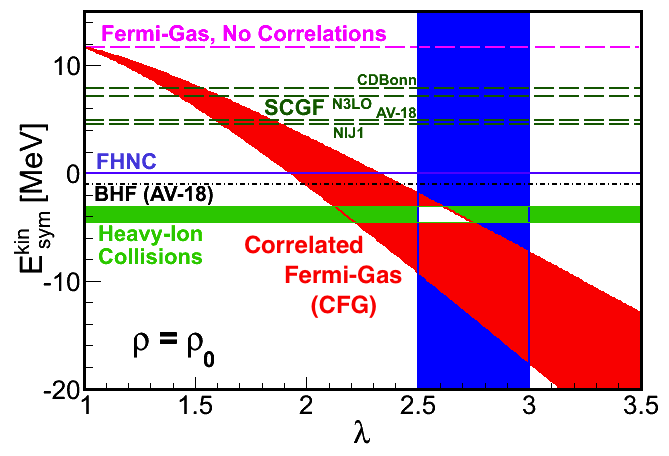}
\caption{(Color online) The per-nucleon kinetic symmetry energy at saturation density, \EsyKZ, calculated using the Correlated Fermi-Gas model (diagonal red band) as a function of $\lambda$, the high-momentum tail cutoff parameter.  The  dashed purple line shows the results of the uncorrelated Fermi Gas model.  The green band shows the results from transport model analyses of Sn+Sn collisions described in the text. Also shown for comparison are the results from microscopic calculations: Brueckner-Hartree-Fock (BHF) \cite{Vid11}, Fermi-Hyper-Netted-Chain (FHNC) \cite{Lov11} and the Self-Consistent Greens Function (SCGF) using the CDBonn, N3LO, Nij1, and AV18 nucleon-nucleon interactions \cite{Car12,Rio14}.}
\label{fig:ekinsym}
\end{figure}

The per nucleon kinetic energy of nuclei and of symmetric nuclear matter can then be calculated from the momentum distribution using 
\begin{equation}
E_{kin} = \frac{4\pi}{(2\pi)^3}\int_0^{\infty} \frac{\hbar^2 k^2}{2m} n(k) k^2 dk .
\end{equation}
Fig.~\ref{fig:ekin} shows the resulting kinetic energy for finite nuclei, calculated within the CFG model using $a_2(A)=5\pm0.3$.  The CFG kinetic energy is much larger than that of the uncorrelated Fermi Gas.  It agrees with
the kinetic energies from many-body nuclear calculations for $^{12}$C, $^{16}$O, $^{40}$Ca, $^{56}$Fe, and $^{208}$Pb~\cite{Ciofi96} and from VMC calculations for $^{12}$C~\cite{Wiringa14}.

Fig.~\ref{fig:ekin_SNM}  shows the  average nucleon kinetic energy for SNM, $E^{kin}_{\mathrm{SNM}}(\rho_0)$ calculated at saturation density, and shown as a function of $\lambda$. The CFG calculation is done using $a_2(\infty) = 7\pm1$ and $R_d=0.64\pm 0.10$, and is compared with the free Fermi gas model and  the predictions of several microscopic models \cite{Vid11,Lov11,Car12,Rio14}. The error band on the CFG results combines estimated uncertainties in $R_d$ and $a_2(\infty)$. The self-consistent Green's function (SCGF) calculations of the kinetic energy of symmetric nuclear matter, $E^{kin}_{SNM}(\rho_0)$ \cite{Car12,Rio14}, agree with our CFG calculation (Fig.~\ref{fig:ekin_SNM}).

Almost all phenomenological and microscopic many-body theories lead to  Equations of State (EOS) of asymmetric nucleonic matter that vary quadratically with the isospin-asymmetry $\delta=(\rho_n-\rho_p)/(\rho_n+\rho_p)$ according to the so-called empirical parabolic law 
$
E(\rho ,\delta )=E(\rho ,\delta =0)+E_{\mathrm{sym}}(\rho )\delta
^{2}+O(\delta ^{4}).  \label{EsymPara}
$
The coefficient of the $\delta ^{4}$ term at $\rho _{0}$ has been found to be less than $1$ MeV \cite{LCK}.
The symmetry energy can thus be calculated equally accurately from either the energy difference between PNM and SNM, i.e., $E_{\mathrm{sym}}(\rho )=E(\rho ,1)-E(\rho ,0)$ or the curvature $E_{\mathrm{sym}}(\rho )=\frac{1}{2}\frac{\partial ^{2}E(\rho,\delta )}{\partial \delta ^{2}}$ at any $\delta$. 

However, it has never been tested whether the empirical parabolic law is valid separately for the kinetic and potential parts of the EOS. While the free Fermi gas  kinetic energy satisfies the parabolic law, models that include SRC may not \cite{XHLi}. To be consistent and compare with the free Fermi gas model and microscopic many-body theories, we will define the kinetic symmetry energy of correlated nucleonic matter as $E^{kin}_{\mathrm{sym}}(\rho )=E^{kin}_{PNM}(\rho)-E^{kin}_{SNM}(\rho).$ We add a SRC correction term to the Fermi Gas symmetry energy to get the full kinetic symmetry energy: 
\begin{equation}\label{CFG}
E^{kin}_{\mathrm{sym}}(\rho )=E_{sym}^{kin}(\rho)\vert_\textrm{FG}-\Delta E_{sym}^{kin}(\rho)
\end{equation}
where the SRC correction term is:
\begin{equation}
\Delta E^{kin}_{sym}\equiv\frac{E^0_F}{\pi^2}c_0\left[\lambda(\frac{\rho}{\rho_0})^{1/3}-\frac{8}{5}(\frac{\rho}{\rho_0})^{2/3}
+\frac{3}{5}\frac{1}{\lambda}(\frac{\rho}{\rho_0})\right].
\label{SRC}
\end{equation}
As one expects, the SRC correction increases with both the height ($c_0=C_\infty/k_F=R_d a_2(\infty)$) and width ($\lambda$) of the high-momentum tail in SNM.

Fig.~\ref{fig:ekinsym} shows the   kinetic symmetry energy, $E^{kin}_{\mathrm{sym}}(\rho_0)$  calculated at saturation density assuming a free Fermi gas model for PNM and shown as a function of $\lambda$.  The error band on the CFG results combines estimated uncertainties in $R_d, a_2(\infty)$ and the amount of SRC in PNM ($<2\%$). Within the uncertainty range of the parameter  $\lambda=2.75\pm 0.25$, $E^{kin}_{\mathrm{sym}}(\rho_0)$ is found to be  between $-2.5$ and $-17.5$ MeV, much less than the free Fermi gas result of $\approx +12.5$ MeV. The microscopic many-body theories yield results that are significantly smaller than the free Fermi gas prediction but significantly larger than our CFG model. Despite the agreement between our CFG model and the SCGF calculations of the kinetic energy of symmetric nuclear matter, $E^{kin}_{SNM}(\rho_0)$ \cite{Car12,Rio14}, the 
 SCGF symmetry energy, $\EsyKZ = E^{kin}_{PNM}(\rho_0) - E^{kin}_{SNM}(\rho_0)$, is significantly larger than our model's. This is because the SCGF calculations include about  10\% correlations in PNM.

To further validate our CFG model, we perform a transport model analysis of nucleon emission data in intermediate energy heavy-ion collisions. The dynamics of heavy-ion collisions around the Fermi energy are sensitive to the density dependence of the nuclear symmetry energy around $\rho_0$ \cite{Bar05,LCK}. Specifically, the ratio of free neutrons to protons emitted in heavy-ion collisions was found to be sensitive to the symmetry energy~\cite{LiBA97a}. This ratio has been measured recently in $^{124}$Sn+ $^{124}$Sn and $^{112}$Sn+ $^{112}$Sn reactions at $E_{beam}/A=50$ and 120 MeV at MSU~\cite{MSU2} with improved precision as compared to earlier measurements \cite{MSU1}. The data are given for the double ratio of neutrons to protons in $^{124}$Sn+ $^{124}$Sn to $^{112}$Sn+ $^{112}$Sn reactions to reduce systematic errors associated with neutron detection.   

Using the Isospin-dependent Boltzmann-Uehling-Uhlenbeck (IBUU) transport model \cite{LCK}, analysis of this double ratio was done by introducing two parameters, $\eta$ and $\gamma$, to describe the potential symmetry energy: 
\begin{equation}
E_{sym}^{pot}(\rho)=[E_{sym}(\rho_0)-\eta E_{sym}^{kin}(\rho_0)\vert_\textrm{FG}] (\rho/\rho_0)^{\gamma}.
\end{equation}
 Without considering the momentum dependence of nuclear potentials, the corresponding symmetry potential is then
\begin{eqnarray}
V^{n/p}_{\rm sym}(\rho,\delta)&=&[E_{sym}(\rho_0)-\eta E_{sym}^{kin}(\rho_0)\vert_\textrm{FG}](\rho/\rho_0)^{\gamma}\nonumber\\
&\times&[\pm 2\delta+(\gamma-1)\delta^2],
\end{eqnarray}
where the $2\delta$ term dominates. The $\pm$ sign is due to the fact that neutrons and protons feel repulsive and attractive symmetry potentials respectively. 

We varied $\eta$ and $\gamma$ on a large 2D fine lattice to minimize the $\chi^2$ between the model calculations and the MSU data at both beam energies.  We then performed a covariance analysis to find the uncertainties of $\eta$ and $\gamma$ corresponding to a $\pm 1\sigma$ error band using the method reviewed recently in Refs. \cite{Dob,Jorge}.  We used an impact parameter of $3$ fm, consistent with that estimated for the data \cite{Cha}. Free nucleons are identified as those with local densities less than $\rho_0/8$ at the time of their final freeze-out from the reaction. Calculations using a phase-space coalescence model lead to similar results within the error band \cite{Li14}.

\begin{figure}[b]
\includegraphics[width=\linewidth]{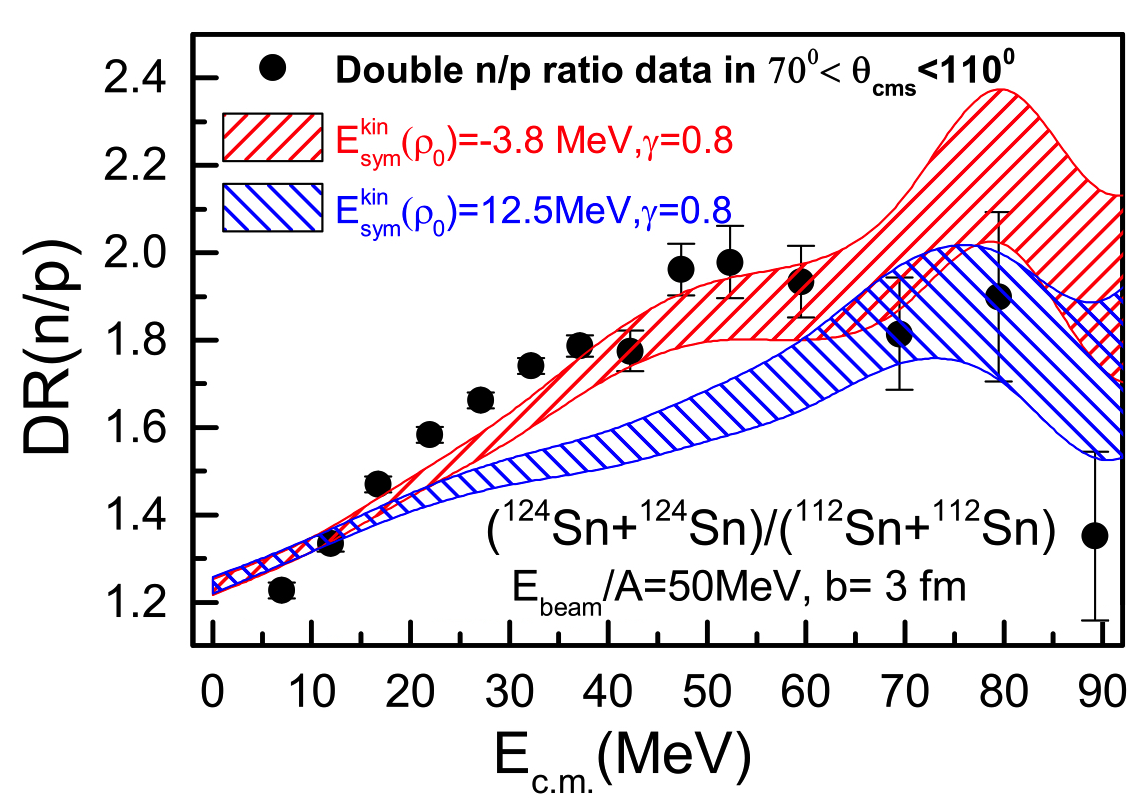}
\caption{(Color online) The calculated double ratio of free neutron/protons in the two reactions in comparison with the MSU data for transversely emitted nucleons in the angular range of $70^0 \le \theta_{cms}\le 110^0$ \cite{MSU2}. The bands represent $1\sigma$ uncertainty of the calculations.   }
\label{fig:SnPnRatio}
\end{figure}

Fig.~\ref{fig:SnPnRatio} shows the double free neutron/proton ratios in the two $124$ and $112$ Sn+Sn reactions at $E_{beam}/A=50$ MeV/nucleon~\cite{MSU2}.
The calculations (red band) shown used the optimized parameters  $\eta_0=-0.30(1\pm 18.53\%)$ that corresponds to $E_{sym}^{kin}(\rho_0)=-(3.8\pm 0.7)$ MeV and $\gamma_0=0.80(1\pm 5.98\%)$ with a $\chi^2_0=8$. This value of $E_{sym}^{kin}(\rho_0)$ was determined without considering the momentum dependence of the symmetry potential known to decrease somewhat the free neutron/proton ratio \cite{Li04}. It thus represents an upper bound on the kinetic symmetry energy used to reproduce the MSU data within the IBUU model. 
For comparison, results with a $\chi^2=21$ using $E_{sym}^{kin}(\rho_0)\vert_\textrm{FG}=12.5$ MeV and $\gamma=0.8$ are also shown. Calculations with $E_{sym}^{kin}(\rho_0)\vert_\textrm{FG}$ and other values of $\gamma$ between 0.4 and 1 leads to even higher $\chi^2$ values.




The value of \EsyKZ{} determined from the IBUU transport analysis of the neutron to proton ratios in Sn+Sn collisions is consistent with that calculated using our CFG model (see Fig.~\ref{fig:ekinsym}).  

\begin{table}[t]
  \caption{Density dependence parameter, $\gamma$, of the potential part of the symmetry energy extracted within the Correlated Fermi Gas (CFG) and Free Fermi Gas (FG) models, assuming a total symmetry energy of $\EsymZ \approx 31$ MeV. Also shown are the value of $\gamma$ and its $1\sigma$ and $2\sigma$ confidence intervals, extracted from analysis of heavy ion collision data~\cite{Tsang09} and neutron stars observations~\cite{Andrew}, assuming a Free FG model. The assumed value of the kinetic symmetry energy at saturation density used in each extraction is also listed.}
\begin{tabular}{| c | c |  c |}
\hline
              & $E_{sym}^{kin}(\rho_0)$ & $\gamma$        \\
              &   [MeV]                            &  $\pm1\sigma(2\sigma)$ \\
\hline
{CFG}                    
              & $-10 \pm 3$                    & $0.25\pm0.05$   \\
\hline
\multirow{4}{*}{FG}                
              & $-10 \pm 3$                    & $0.58\pm0.05$   \\
              & $   0   $                           & $0.55\pm0.06$   \\
              & $12.5 $                           & $0.48\pm0.10$   \\
              & $17.0 $                           & $0.41\pm0.13$   \\
\hline
Tsang et al.~\cite{Tsang09}               
              & $12.5$                            & $0.7^{+0.1(0.35)}_{-0.2(0.3)}$   \\
\hline
Steiner et al.~\cite{Andrew}               
              & $17.0$                            &  $0.3^{+0.1(0.5)}_{-0.1}$    \\
\hline
\end{tabular}
\label{tab:1}
\end{table}

We now turn to extracting the total symmetry energy at supra-nuclear densities and the density dependence of its potential part using the CFG model.
We use the general form of the total symmetry energy given by Eq.~\ref{eq:a}, with the CFG corrections to the kinetic energy term given by Eq.~\ref{CFG} and~\ref{SRC}. As detailed above, by comparing the CFG model results to the known values of the total symmetry energy ($\Esym=31.0\pm1(1\sigma)$ MeV~\cite{Lat13}) we can extract the value of the potential part of the symmetry energy at saturation density: $E_{sym}^{pot}(\rho_0)=E_{sym}(\rho_0)-E_{sym}^{kin}(\rho_0)$. Simillarly, using the known density dependence of the total symmetry energy at saturation density ($L=50\pm5(1\sigma)$ MeV~\cite{Lat13}) we can extract the density dependence of the potential part of the symmetry energy:
\begin{equation}
\gamma = \frac{\frac{1}{3}L-\frac{dE_{sym}^{kin}(\rho)}{d\rho}|_{\rho_0}}{E_{sym}(\rho_0)-E_{sym}^{kin}(\rho_0)}.
\end{equation} 
Our results are summarized in Table~\ref{tab:1} where we list the value of $\gamma$ extracted using the CFG model. This is compared with free fermi gas model results (i.e. $\alpha=2/3$) assuming different values for the kinetic symmetry energy (i.e. $\EsyKZ=-10,0,12.5,17$ MeV), and with recent analyses of heavy ion collisions~\cite{Tsang09} and neutron star data~\cite{Andrew} which also assume a free fermi gas model (i.e. $\alpha=2/3$). As can be seen, even within the FG model, the value of $\gamma$ varies significantly depending on the  value of the kinetic symmetry energy. Furthermore, CFG and FG results for the same kinetic symmetry energy also differ due to the density dependence of the SRC correction term (eq.~\ref{SRC}). The value of $\gamma$ obtained from the neutron star analysis of Ref.~\cite{Andrew} is very similar to that of the CFG model.

\begin{figure}[t]
\includegraphics[width=\linewidth]{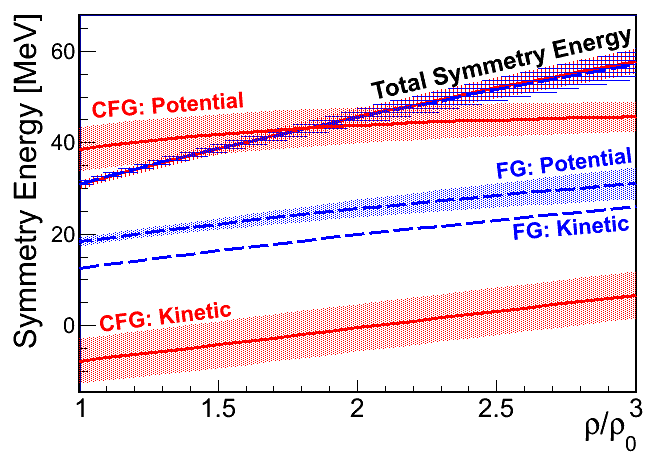}
\caption{(color online) The density dependence of the kinetic, potential and total symmetry energy extracted  using the CFG and  FG models.  See text for details.}
\label{fig:Esym}
\end{figure}

Fig.~\ref{fig:Esym} shows the density dependence of  the kinetic, potential  and  total symmetry energy obtained using both the CFG and FG models. While the two models differ significantly in the values and density dependences of their kinetic and potential parts, their total symmetry energies are almost identical.

To summarize, we provide an analytical expression for a  kinetic symmetry energy of correlated  nucleonic matter at $\rho=\rho_0$, using the dominance of  short-range correlated neutron-proton pairs at high momentum observed in electron scattering data.
Our  model yields  $\EsyKZ=-10\pm7.5$ MeV, significantly lower than $\EsyKZ=+12.5$ MeV of the widely-used free Fermi gas model.  This result is consistent with our analysis of recent data on the free proton-to-neutron ratios measured in intermediate energy nucleus-nucleus collisions as well as with microscopic many-body calculations, and previous phenomenological extractions.
We also extract the density dependence of \EsyP{} and \Esym{} from
our model of \EsyK{} together with  the value of the total symmetry energy and its density dependence at saturation density.  While the total symmetry energy exacted using different models is consistent, its separation into kinetic and potential parts is not.

We thank F.J. Fattoyev, X. H. Li, W.G. Newton, Z.Z Shi, A. Gal, M. M. Sargsian, G. A. Miller, D. W. Higinbotham, M. Strikman, and  L. Frankfurt for helpful discussions. 
O. Hen and E. Piasetzky are  supported by the Israel Science Foundation. B.A. Li is supported in part by  the US National Science Foundation under Grant No. PHY-1068022, US National Aeronautics and Space Administration under Grant No. NNX11AC41G issued through the Science Mission Directorate, the CUSTIPEN (China-U.S. Theory Institute for Physics with Exotic Nuclei) under DOE grant number DE-FG02-13ER42025 and the National Natural Science Foundation of China under Grant No. 11320101004. W.J. Guo is supported by the National Natural Science Foundation of China (10905041) and the China Scholarship Council Foundation (201208310156). L.B. Weinstein is supported by the US Department of Energy under Grants DE-SC00006801 and DE-FG02-96ER40960.


\begin{thebibliography}{0}
\expandafter\ifx\csname natexlab\endcsname\relax\def\natexlab#1{#1}\fi
\expandafter\ifx\csname bibnamefont\endcsname\relax
  \def\bibnamefont#1{#1}\fi
\expandafter\ifx\csname bibfnamefont\endcsname\relax
  \def\bibfnamefont#1{#1}\fi
\expandafter\ifx\csname citenamefont\endcsname\relax
  \def\citenamefont#1{#1}\fi
\expandafter\ifx\csname url\endcsname\relax
  \def\url#1{\texttt{#1}}\fi
\expandafter\ifx\csname urlprefix\endcsname\relax\def\urlprefix{URL }\fi
\providecommand{\bibinfo}[2]{#2}
\providecommand{\eprint}[2][]{\url{#2}}

\end{thebibliography}


\begin{thebibliography}{99}

  \bibitem{Lat91} J.M. Lattimer, C.J. Pethick, M. Prakash, P. Haensel,
Phys. Rev. Lett. 66, 2701 (1991). 

 \bibitem{Pra88b} M. Prakash, T.L. Ainsworth, J.M. Lattimer, Phys. Rev.
Lett. 61, 2518 (1988). 

\bibitem{Nikolov11} N. Nikolov, N. Schunck, W. Nazarewicz, M. Bender, and J. Pei, Phys. Rev. C 83, 034305 (2011).

\bibitem{LiBA97a} B.A. Li, C.M. Ko, Z.Z. Ren, Phys. Rev. Lett. 78, 1644 (1997).

\bibitem{Che05a} L.W. Chen, C.M. Ko, B.A. Li, Phys. Rev. Lett. 94, 032701 (2005).

\bibitem{Tsang09} M.B.Tsang et al.,  Phys. Rev. Lett. 102, 122701,(2009).   
 
\bibitem{FRIB}  A. B. Balantekin et al.,  Modern Physics Letters A {\bf 29}, 30010 (2014). 
  
\bibitem{NASA} C. Kouveliotou et al., Enduring Quests-Daring Visions (NASA Astrophysics in the
Next Three Decades). arXiv: 1401.3741, January 2014.

\bibitem{Andrew1} A.W. Steiner, M. Prakash, J.M. Lattimer, P.J. Ellis,
Phys. Rep. {\bf 411}, 325 (2005).

\bibitem{LiBA06a} B. A. Li and A. W. Steiner, Phys. Lett. B \textbf{642}, 436 (2006).

\bibitem{Will} W.G. Newton, M. Gearheart and B.A. Li, 
APJ Supplementary Series 204, 9 (2013). 

\bibitem{Andrew} A. W. Steiner, J. M. Lattimer, and E. F. Brown, Astrophys.
J. 722, 33 (2010).

\bibitem{Lynch09} W.G. Lynch {\it et al.}, Prog. Nucl. Part. Phys. {\bf 62}, 427 (2009).

\bibitem{Trau12} W. Trautmann and H. H. Wolter, Int. J. Mod. Phys. E {\bf 21}, 1230003 (2012).

\bibitem{Tsang12} M. B. Tsang, et al., Phys. Rev. C {\bf 86}, 015803 (2012).

\bibitem{EPJA} B.A. Li, A. Ramos, G. Verde, and I. Vida\~na, eds., "Topical issue on nuclear symmetry energy", Eur. Phys. J. A {\bf 50}, No. 2, (2014).

 \bibitem{Hor14} C.J. Horowitz et al., J. of Phys. G {\bf 41}, 093001 (2014). 
  
 \bibitem{Jim13} J. M. Lattimer, Annu. Rev. Nucl. Part. Sci. \textbf{62}, 485 (2012).
 
  \bibitem{Lat13} J.M. Lattimer and Y. Lim,
Astrophys. J. 771, 51 (2013). 

 \bibitem{LiBA13} B.A. Li, X. Han, Phys. Lett. B \textbf{727}, 276 (2013).


\bibitem{Gandolfi14} S. Gandolfi, J. Carlson, S. Reddy, A. W. Steiner, and R. B. Wiringa, 
 Eur. Phys. J A {\bf 50} (2014) 10.

\bibitem{Hebeler14} K. Hebeler and A. Schwenk,  Eur. Phys. J A {\bf 50} (2014) 11.


\bibitem{Steiner12} A. W. Steiner, S. Gandolfi
 Phys. Rev. Lett. {\bf 108}, 081102 (2012).

\bibitem{Hebeler13} K. Hebeler, J. M. Lattimer, C. J. Pethick, A. Schwenk
 Astrophys.J. {\bf 773} (2013) 11.

\bibitem{Gezerlis13} A. Gezerlis, I. Tews, E. Epelbaum, S. Gandolfi, K. Hebeler, A. Nogga, A. Schwenk
 Phys. Rev. Lett. {\bf 111}, 032501 (2013).

\bibitem{Li14}B.A. Li, W.J. Guo and Z.Z. Shi, arXiv:1408.6415  (2014). 

 
\bibitem{XuLi} C. Xu, A. Li, B.A. Li, J. of Phys: Conference Series 420, 012190 (2013).

\bibitem{Vid11} I. Vidana, A. Polls, C. Providencia, Phys Rev C 84, 062801(R) (2011).

\bibitem{Lov11} A. Lovato, O. Benhar, S. Fantoni, A. Yu. Illarionov, and K. E. Schmidt, Phys. Rev. C 83, 054003 (2011).

\bibitem{Car12} A. Carbone, A. Polls, A. Rios, Eur. Phys. Lett. 97, 22001 (2012).

\bibitem{Rio14} A. Rios, A. Polls, W. H. Dickhoff, Phys. Rev. C {\bf 89}, 044303 (2014).

\bibitem{Bar05} V. Baran, M. Colonna, V. Greco, and M. Di Toro, Phys. Rep.
\textbf{410}, 335 (2005).

\bibitem{LCK} B. A. Li, L. W. Chen, and C. M. Ko, Phys. Rep. \textbf{464}, 113 (2008).

\bibitem{Egi03} K. Egiyan, et al. (The CLAS Collaboration), Phys. Rev. C 68, 014313 (2003).

\bibitem{Egi06} K. Egiyan, et al. (The CLAS Collaboration), Phys. Rev. Lett. 96, 082501 (2006).

\bibitem{Fom12} N. Fomin, et al. (The Hall C Collaboration), Phys. Rev. Lett. 108, 092502 (2012).

\bibitem{Hen14a} O. Hen et al. (The CLAS Collaboration), Science 346, 614 (2014).

\bibitem{sub} R. Subedi, et al. (The Hall A Collaboration), \textit{Science} \textbf{320},  1476 (2008).

\bibitem{MSU1}M.A. Famiano et al., Phys. Rev. Lett. 97, 052701 (2006).

\bibitem{MSU2} D.S. Coupland et al., arXiv:1406.4546

\bibitem{Wiringa14} R.B. Wiringa, R. Schiavilla, Steven C. Pieper, and J. Carlson, Phys. Rev. C {\bf 89}, 024305 (2014).

\bibitem{Ciofi96} C. Ciofi degli Atti and S. Simula, Phys. Rev. C 53, 1689 (1996). 


\bibitem{Bethe71} H.A. Bethe, \textit{Ann. Rev. Nucl. Part. Sci.} \textbf{21}, 93 (1971).

\bibitem{ant} A.N. Antonov, P.E. Hodgson and I.Z. Petkov, \textit{Nucleon Momentum
and Density Distributions in Nuclei} (Clarendon Press, Oxford, 1988).

\bibitem{fan} S. Fantoni and V.R. Pandharipande, \textit{Nucl. Phys. A} \textbf{427}, 473 (1984).

\bibitem{Pie92} S.C. Pieper, R.B. Wiringa and V.R. Pandharipande, \textit{Phys. Rev. C} {\bf 46},1741 (1992).

\bibitem{Day89} D.B. Day {\it et al.}, Phys. Rev. C {\bf 40}, 1011 (1989).

\bibitem{Cio91} C. Ciofi degli Atti, E. Pace, and G. Salme, Phys Rev C {\bf 43}, 1155 (1991).

\bibitem{Sar11} Michael McGauley and Misak M. Sargsian, arXiv 1102.3973.

\bibitem{Pia13} E. Piasetzky, O. Hen, and L. B. Weinstein, AIP Conf. Proc. 1560, 355 (2013).


\bibitem{pia} E. Piasetzky et al., \textit{Phys. Rev. Lett.} \textbf{97}, 162504 (2006).

\bibitem{Tan03} A. Tang et al., \textit{Phys. Rev. Lett.} {\bf 90}, 042301 (2003).

\bibitem{Bag10} H. Baghdasaryan, et al. (The CLAS Collaboration), \textit{Phys. Rev. Lett.} {\bf 105}, 222501 (2010).


\bibitem{fra} L. Frankfurt, M. Sargsian and M. Strikman, \textit{Int. Jour. Mod. Phys. A} \textbf{23}, 2991 (2008).

\bibitem{Arr12} J. Arrington, D.W. Higinbotham, G. Rosner G and M. Sargsian,
Prog. Part. Nucl. Phys. 67, 898 (2012).

\bibitem{Pas02} I. Passchier et al., Phys. Rev. Lett. 88 , 102302 (2002).

\bibitem{Hen14b} O. Hen, L. B. Weinstein, E. Piasetzky, G. A. Miller, M. M. Sargsian, Y. Sagi, arXiv:1407.8175.




\bibitem{XHLi}X.H. Li et al., arXiv:1403.5577 (2014).

\bibitem{Dob} J. Dobaczewski, W. Nazarewicz, P.-G. Reinhard,
J. Phys. G: Nucl. Part. Phys. 41, 074001 (2014).

\bibitem{Jorge} J. Piekarewicz, Wei-Chia Chen and F.J. Fattoyev, arXiv:1407.0911

\bibitem{Cha} Z.Chajecki and Betty Tsang, private communications (2014).

\bibitem{Li04} B.A. Li, C. B. Das, S. Das Gupta, C. Gale, Nucl. Phys. A 735, 563 (2004). 
  




\end{thebibliography}
\end{document}